# Brain tumor MRI Classification using a Novel Deep Residual and Regional CNN

Mirza Mumtaz Zahoor [1,2] , Saddam Hussain Khan [3*]

1. Department of Computer & Information Sciences(DCIS), Pakistan Institute of Engineering and Applied Sciences (PIEAS), Islamabad 45650, Pakistan; mumtazzahoor5@gmail.com (M.M.Z.);
2. Faculty of Computer Sciences, Ibadat International University, Islamabad, Pakistan
3. Department of Computer Systems Engineering, University of Engineering and Applied Science (UEAS), Swat, Pakistan; saddamhkhan@ueas.edu.pk (S.H.K.)

Corresponding author: (e-mail: saddamhkhan@ueas.edu.pk).

## Abstract:

Brain tumor classification is crucial for clinical analysis and an effective treatment plan to cure patients. Deep learning models help radiologists to accurately and efficiently analyze tumors without manual intervention. However, brain tumor analysis is challenging because of its complex structure, texture, size, location, and appearance. Therefore, a novel deep residual and regional-based Res-BRNet Convolutional Neural Network (CNN) is developed for effective brain tumor Magnetic Resonance Imaging (MRI) classification. The developed Res-BRNet employed Regional and boundary-based operations in a systematic order within the modified spatial and residual blocks. Spatial blocks extract the brain tumor's homogeneity, heterogeneity, and boundary with patterns and edge-related features. Additionally, the residual blocks significantly capture local and global texture variations of brain tumors. The efficiency of the developed Res-BRNet is evaluated on a standard dataset; collected from Kaggle and Figshare containing various tumor categories, including meningioma, glioma, pituitary, and healthy images. Experiments prove that the developed Res-BRNet outperforms the standard CNN models and attained excellent performances (accuracy: 98.22%, sensitivity: 0.9811, F1-score: 0.9841, and precision: 0.9822) on challenging datasets. Additionally, the performance of the proposed Res-BRNet indicates a strong potential for medical image-based disease analyses.

# 1. INTRODUCTION

The human brain is among the body's most complicated and imperative organs, governing the neurological system. The most deadly brain tumor is caused by erratic and out-of-control cell growth in the brain [1]. Patient Survival depends on the type of glioma type; lowgrade gliomas have survival rates of 5 years as high as 80%, whereas survival rates of 5 years are under 5% for high-grade gliomas [2]. Timely brain tumor recognition and categorization is an imperative research topic in the clinical imaging domain, and it assists in choosing the most suitable treatment plan for a patient's life-saving [3].

Several screening methods, either invasive or non-invasive, are employed to identify tumors in the human brain [4]. Magnetic resonance imaging (MRI) is a preferable, less harmful scanning modality since it provides rich information about the location of brain tumors, shape, and size in medical images (MI) and is generally considered quicker, cheaper, and safer [5]. Manual assessment of brain MR scans is challenging for radiologists to identify and categorize brain tumors from MIs. A computer-aided diagnosis (CADx) is required to reduce the burden and assist radiologists or doctors with MI assessment.

Many research areas are being explored in medical image analysis. It includes medical imaging domains like identification, detection, and segmentation [6]–[11]. Traditional ML approaches comprise numerous steps, pre-processing, feature extraction and selection, and classification. More discriminative feature acquisition is essential, as classification accuracy relies on obtained features.

In conclusion, the conventional ML techniques have two key challenges in the feature extraction step. One, it solely emphasizes low- or high-level attributes. Secondly, standard ML techniques rely on hand-crafted features that require significant prior knowledge, such as the position of the tumour in a medical scan, However, there is a considerable risk of human error. Designing an effective system to incorporate high- and low-level features with no human intervention is crucial. As brain tumor datasets are being expanded, there is a need for technological improvements in feature extraction focusing on confined and imbalanced MR imaging data sets of brain abnormalities and other irregularities of the human organs[12] [13].

Recently, deep learning (DL) methods have frequently been employed for brain MRI categorization [14]. While feature mining and classification were integrated into self-learning, deep learning methods do not necessitate a manual process for feature extraction. The DL approach requires a dataset, and minimal pre-processing is required for selecting salient features in a self-learning way [15]. MR imaging categorization faces a significant challenge in diminishing the semantic space among high-level spatial details observed by a human assessor and low-level acquired using the imagery mechanism. One of the well-known neural network models, convolutional neural networks (CNNs), specially designed for images, is utilized for feature extraction to acquire the important characteristics to categorize and minimize the semantic gap.

Recently, in many studies, CNNs have been widely employed to classify brain MRI and validated on a different dataset of brain tumors [16]–[20]. A deep CNN-based model was proposed in [21] for brain MRI images categorization into distinct classes. The authors used brain MRI images from a publicly available dataset to prevent model ambiguity. The suggested model has a classification accuracy of 91.4%. Deepak and Ameer [22] employed a pre-train deep CNN, GoogLeNet, to extract key attributes using brain MR images and classify tumors into three classes with 98% accuracy. Ahmet and Muhammad [23] categorized brain MR images using various CNN models and attained satisfactory accuracy. They modify a pre-trained ResNet-50 DCNN by excluding the final five layers and introducing additional eight layers. The model achieved the highest among all pre-trained models accuracy of 97.2 %. Sultan et al. [24] suggested a CNN-based deep learning model utilizing two publicly accessible datasets have 3064 (glioma, meningioma, and pituitary tumors) and 516 (Grade II, Grade III, and Grade IV) brain medical scans. The proposed method has the best accuracy of 96.13 % and 98.7 %. Khwaldeh et al. [25] used several CNNs to classify brain MRI images and achieved good results. Using modified pre-trained Alexnet CNN, they achieved a higher accuracy of 97.2 %. Khan, M.A. et al. [26] developed a multi-model-based technique to differentiate brain tumors with DL. The presented system includes many stages. They used partial least squares (PLS) to concatenate the features and ELM for classification. Their methodology stated improvement of 97.8%, 96.9%, and 92.5% on BraTs-2015, BraTs-2017, and BraTs-2018, respectively. Özyurt et al. [27] presented a technique for detecting brain tumors. They began with MRI tumor image segmentation with the NS-EMFSE algorithm. They obtained features from the segmented image using AlexNet and then using the SVM, they detected and classified brain tumor images as benign or malignant with 95.62 % accuracy. However, most of these models are assessed on small-scale datasets due to the inaccessibility of the data repositories. Likewise, the majority of earlier research was based on pre-trained CNN models, which were developed generally for a dataset of natural images. Pretrained models are customized for the brain tumor task without designing them to distinguish brain tumor patterns. Thus it limits the use of pre-trained CNN models for brain tumor diagnosis.

In this study, a new deep CNN-based brain tumor classification scheme is developed for MRI image categorizing. A novel CNN architecture, Res-BRNet, is suggested for brain tumor classification. Performance assessment is performed using standard measures like sensitivity, precision, F1-score, accuracy, and AUC-PR/ROC. Moreover, we have generated a large dataset by collecting brain MRI images of three tumor types and normal brain images from publicly accessible sources. The prediction ability of the developed approach is assessed on the test dataset and assessed with a comparison of numerous existing DCNNs, and also proposed technique's idea is compared with baseline approaches. The proposed work has the following contributions:

- A new deep residual and regional CNN architecture, Res-BRNet, is developed for brain tumor classification.
- The developed Res-BRNet employed Regional and boundary-based operations in a systematic order within the modified spatial and residual blocks to exploit spatial correlation information and textural variations from brain tumor MRIs.
- The systematic integration of residual and spatial blocks within the proposed Res-BRNet CNN improves discriminative capability and generalization. Moreover, spatial blocks extract homogeneity and boundary-defined features at the abstract level. Furthermore, residual blocks at the target level effectively learn local and global texture variations of different brain tumors.
- The proposed brain tumor Res-BRNet classification CNN significantly reduces false positives and negatives compared to several existing CNN architectures.

The rest of the manuscript is organized as follows: Section 2: incorporates the proposed methodology. The results and discussion are described in section 3, and the last Section 4: concludes the entire paper.

## 2. Methodology

In this work, a new deep residual and regional CNN architecture is designed for automated brain tumor classification from MRI images. The discriminating ability of the proposed classification method is empirically assessed using several standard performance measures, and results are evaluated by comparing them with existing DCNNs. A better generalization is achieved by augmenting the training samples in the experimental setup. The general framework of the developed brain tumor classification technique is shown in Fig. 1.

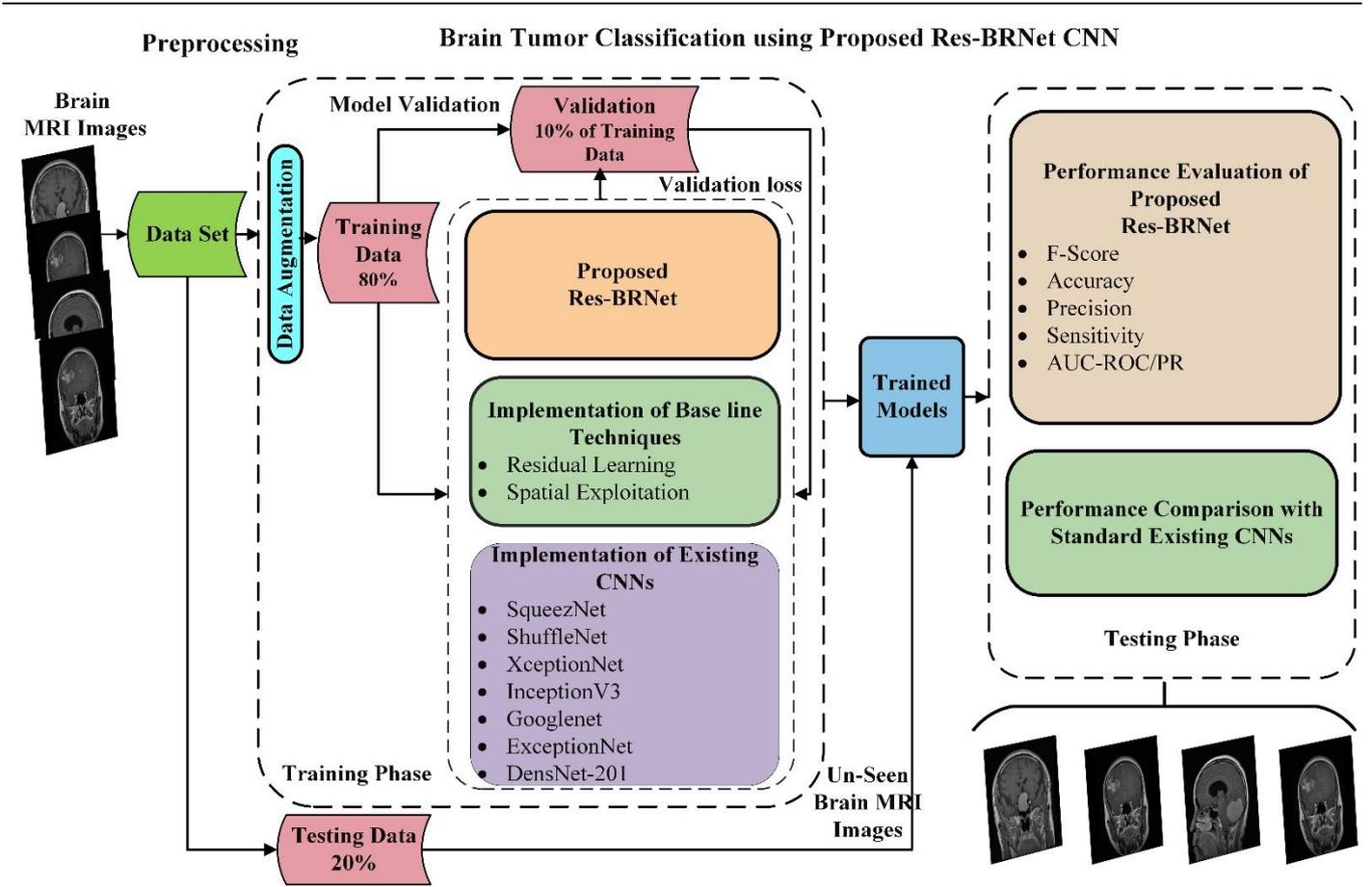

**Figure 1.** The detailed design of the proposed brain tumor MRI images classification technique

## 2.1. Data Set

In this work, we collected a dataset containing MRI images of healthy individuals and three diverse types of brain tumors. MRI scans of four classes are gathered from open-source Kaggle repositories [28], Br35H [29], and figshare [30]. For this experimental setup, we collected 2044 brain normal, 2352 glioma_tumor, 1645 meningioma_tumor, and 1831 pituitary_tumor MRI images from these repositories; hence, in nature, the acquired dataset is unbalanced. Each image was resized to 227 × 227 pixels. Some of the four classes' images are displayed in Fig. 2.

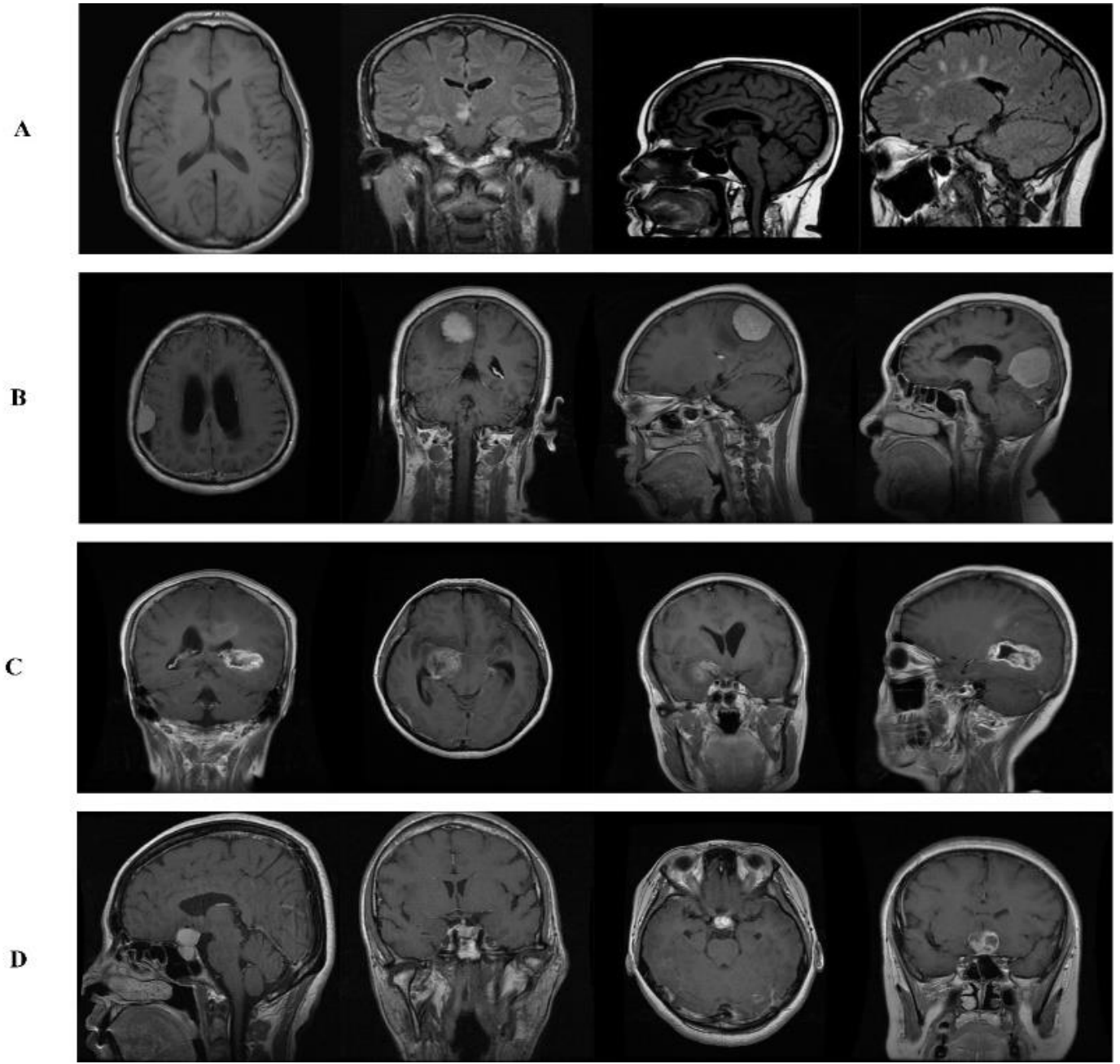

**Figure 2.** Example MRI images of normal and different types of tumors (A) Normal, (B) Glioma_tumor, (C) Meningioma_tumor, and (D) Pituitary_tumor

## 2.1. Data Augmentation

On a small volume of data, deep learning models generally overfit. Thus, a significant amount of data is usually required to train deep CNNs and to provide better generalizability. Data augmentation is generally employed for increasing the original dataset's samples [6], [31], [32]. In this experiment, random rotation (0 - 360 degrees), scaling (0.5 - 1), sharing (± 0.05), and image reflecting (±1 range) are used to augment the data set. These augmentation techniques are used to strengthen the model's generalization.

## 2.2. The developed deep Res-BRNet-based categorization

In this work, we exploit the learning capability of deep CNN to acquire the tumor's distinctive patterns in brain MRI scans. The strong potential of deep CNN for learning specific features and patterns from images inspires us to employ them for classification and recognition tasks. Because of their effective learning capability, CNNs are largely employed for feature extraction and classification. In this proposed work, we designed a novel residual and regional CNN architecture-using boundary and region-based operations to classify tumor-specific abnormalities in brain MRI images and named it Res-BRNet. The proposed model is trained in an end-to-end way to learn the tumor-related patterns from MRI scans. The last fully connected layers, followed by softmax-based operation of the proposed deep CNNs, are used for the final classifications. The details of Res-BRNet are depicted in the section given below.

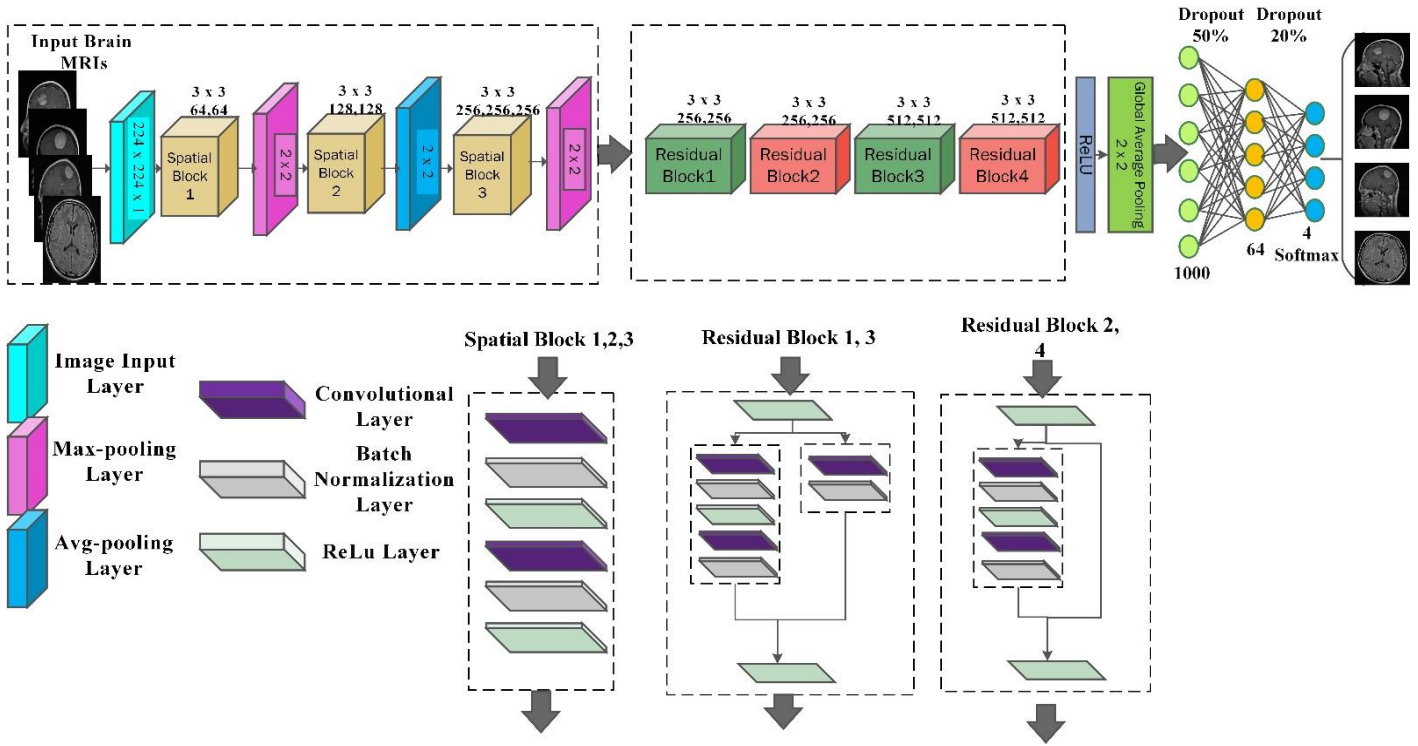

**Figure 3.** Blockwise details of the proposed Res-BRNet

## 2.2.1. Structural details of the developed Res-BRNet

The architecture-level details of the developed Res-BRNet are inspired by standard image processing techniques [33][34]. It is developed to explore the hidden insights in MRI images. In this context, the region and boundary-based operators with convolution operators are optimized in the proposed architecture to attain the brain tumor patterns excellently. In this work, we exploited spatial and residual blocks [35]–[37] as baselines to justify the advantages of boundary uniformity and boundary-related features for obtaining the tumor patterns using CNNs.

As illustrated in Fig. 4 in a spatial block, input x is fed into the operation block, and all operators are applied sequentially on input and at the output operation block, we get T (x)= $f_{conv.}(x)$ as sown in Eq. 1. Difference between a feed-forward spatial block and residual learning is that in residual block skip connections from input x to the output of the encoding block and add up with the output of the encoding block $f_{conv.}(x)$. At the output of the residual block, we get T (x)= $f_{conv.}(x)$+x as shown in Eq. 2 and 3. As compared to spatial block, residual learning facilitates the model to capture minor textural and contrast variations and also facilitates overcoming the vanishing gradient problem, as well as improves the learned features maps and model's convergence.

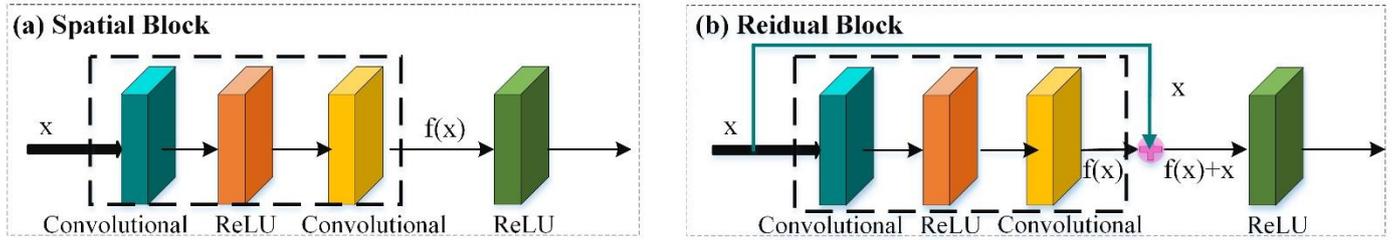

**Figure 4**. The difference in the process of (a) plain and (b) residual block

$$T(x) = f_{conv.}(x) \qquad (1)$$
$$T(x) = f_{conv.}(x) + x \qquad (2)$$
$$f_{conv.}(x) = T(x) - x \qquad (3)$$

The proposed Res-BRNet comprises three spatial blocks at the start, and four residual blocks are used after them. Every spatial block contains a single convolution layer (Eq.4), batch normalization, and ReLU. The convolution-layer exploits tumor-relating patterns, while ReLU performs as an activation function. To learn region homogeneousness and boundary-related attributes of brain tumors, a max- or average-pooling operation is applied at the end of each spatial block, as shown in Eq. (5) and (6).

Fig. 3 illustrates the architecture of the developed Res-BRNet. Fully connected (FC) layers stated in Eq. (7) are applied in the designed architecture to attain particular features for classification. Dropout layers are used with FC layers to minimize the risks of overfitting.

$$Z_{m,n} = \sum_{u=1}^{r}\sum_{v=1}^{s} Z_{m+u-1,\ n+v-1}\ k_{a,b} \qquad (4)$$

$$Z_{m,n}^{Avg} = \frac{1}{T^2} \sum_{u=1}^{t} \sum_{v=1}^{t} Z_{m+u-1,\ n+v-1} \tag{5}$$

$$Z_{m,n}^{Max} = \text{Max}_{u=1\ldots t, v=1\ldots t} Z_{m+u-1,\ n+v-1} \tag{6}$$

$$Q = \sum_{b}^{B} \sum_{c}^{C} W_d Z_c \tag{7}$$

$Z$ illustrates the source feature map of size $M \times N$, and the filter with size $r \times s$ is defined by k in the convolutional operator used in Eq. (4). The output map of features is shown by $Z$. m and n have begun from 1 to $(M - r + 1)$ and $(N - s + 1)$, accordingly. As shown in Eqs. (4)–(6), we regulate the $Z$avg and $Z$max methods, denoted by $Z$ Avg and $Z$ Max, similarly. In Eqs. (5) and (6), t indicates the average- and max-window dimensions. In Eq. (7), the dense layer outcome is stated by Q, which employs global operation on $Z_c$. FC-layer neurons are presented by $W_d$ and preserves essential features for the analysis.

### 2.2.2. Benefits of the proposed Res-BRNet for image contents analysis

Brain MRI scans reveal complex patterns with different intensity levels in distinct regions. Regional smoothness, textural differences, and edges make the basic structure of these patterns. In this study, the developed model is significantly improved by combining the convolutional operator (Eq. 4)), enhancing the region-homogeneity, and boundary-based operations (Eqs. (5) and ((6)), respectively, to differentiate the healthy instances from the tumor-affected MRI scans. In contrast to the developed model, the majority of existing CNN designs employ different convolutional combinations with simply one type of pooling layer to capture invariant features [36], [38]–[42]. The following are the significance of applying the proposed idea in CNN:

- The developed residual and regional CNN architecture aimed to dynamically exploit image smoothness and sharpness, and it may efficiently optimize the level of smoothness and sharpening in harmony with the spatial features of an image.
- Implementing the spatial block (Eq. (1)) with residual learning improves the overall detection ability of the model by acquiring textural features along with spatial correlation from MRI images.
- The systematic use of boundary and regional operations within spatial blocks helps to enhance the region homogeneity of various regions. Using average-pooling (Eq. (5)), the region operator helps to smooth the regional variations and eliminates noise caused by distortions captured during MRI imaging. On the other hand, Res-BRNet is aided by boundary operators to acquire discriminative local features with the max-pooling operation (Eq. (6)).

- Residual blocks (Eq. (2)) facilitate the model to capture textural and minor contrast variations and lead to overcoming the vanishing gradient problem, which is generally produced in very deep architectures.
- Down-sampling is also performed during pooling operations, which increases the model's robustness to small changes in the input image.

### 2.3. Employment of Existing CNNs

Competitive assessment is performed by implementing several existing deep CNN models, including SqueezeNet, ShuffleNet, VGG-16, Xception, ResNet-18, GoogleNet, Inception-V3, and DenseNet-201 [36], [39], [40], [42]–[47]. Several researchers applied these CNNs to classify MRI images and they have been widely used for many image recognition tasks. Although these models' block architecture and design changed, they all employed a single pooling operation along the network or changed this for a stridden convolution operation to control complexity. To fine-tune these CNNs for brain tumor classification, we added FC and a classification layer and employed them in an end-to-end manner.

### 2.4. Implementation details

A brain MRI dataset was split into two sets, 80% train set, and 20% test set. Furthermore, the train set was divided into train and validation data to select optimized parameters. 'RMSprop' [48] was employed for optimization with a 'SquaredGradientDecayFactor' of 0.95 throughout the training of CNNs. The learning rate was initially set to 0.0001 with the "LearnRateDropFactor" to be 0.4 and 40 epochs. A small-batch-based technique is used to train models on a batch size of 16 for every epoch. As an activation function, softmax was used, and cross-entropy loss has been reduced for all of the deep CNNs optimizations. This simulation was conducted using MATLAB 2020b. During MATLAB-based simulations, a 2.90-GHz Dell, Core I i7-7500 CPU, and a Nvidia® GTX 1060 Tesla graphics card with CUDA support have been used.

## 3. Results and discussion

This study suggests a deep CNN-based system for identifying brain tumor patients using MRI images. We perform two different experiments for empirical evaluation of the developed technique. We initially explore the impacts of using simultaneously average- and max-pooling in spatial blocks of Res-BRNet.

Secondly, a general assessment of brain tumor classification is carried out by comparing performances with well-known existing deep CNN models.

## 3.1. Performance metrics

The efficiency of the developed model is assessed using several standard evaluation measures. These measures include precision [49], sensitivity [50], accuracy [51], F1-score [52], PR, and ROC curves [53]. TP defined truly positive predictions, TN as truly negative predictions, FP as incorrectly positive predictions, and FN for incorrectly negative predictions. In (Eq. (8)), accuracy is defined, it calculates the total number of accurate selections. Accordingly, Sensitivity is in (Eq. (9)), precision is denoted in (Eq. (10)), and F1-score is defined in (Eq. (11)).

$$\text{Accuracy} = \left(\frac{TP + TN}{TN + TP + FN + FP}\right) \times 100 \tag{8}$$

$$\text{Sensitivity} = \frac{TP}{TP + FN} \tag{9}$$

$$\text{Precision} = \frac{TN}{TN + FP} \tag{10}$$

$$F1 - \text{Score} = \frac{2 \times (P \times R)}{P + R} \tag{11}$$

## 3.2. Efficiency analysis of the proposed Res-BRNet

In a comprehensive experimental investigation, the proficiency of the developed Res-BRNet is assessed with well-known CNNs on unseen test data using Accuracy, F1-score, Sensitivity, Precision, ROC, and PR-AUC. In contrast to accuracy, F1-score tends to give more weight to precision and sensitivity. Training-loss and accuracy chart for Res-BRNet is presented in Fig. 6. The developed CNN converges smoothly and quickly to achieve its optimal value, as seen in Fig. 6. The proposed Res-BRNet model correctly classified 1553 samples of three brain tumors and normal instances. Likewise, the proposed Res-BRNet performs similarly by correctly identifying 463 gliomas, 321 meningiomas, 365 pituitary, and 404 normal individuals correspondingly. It is observed that a change in the region and boundary arrangements, as illustrated in Fig. 3, improves the overall performance. Fig. 5 displays some of the brain MRI images those Res-BRNet misclassifies. Low contrast, irregular sample patterns, and varying illumination variations are probable reasons for misclassification. The generalization and the robustness enhancement of test samples are achieved by using several data-augmentation strategies while train-developed CNNs.

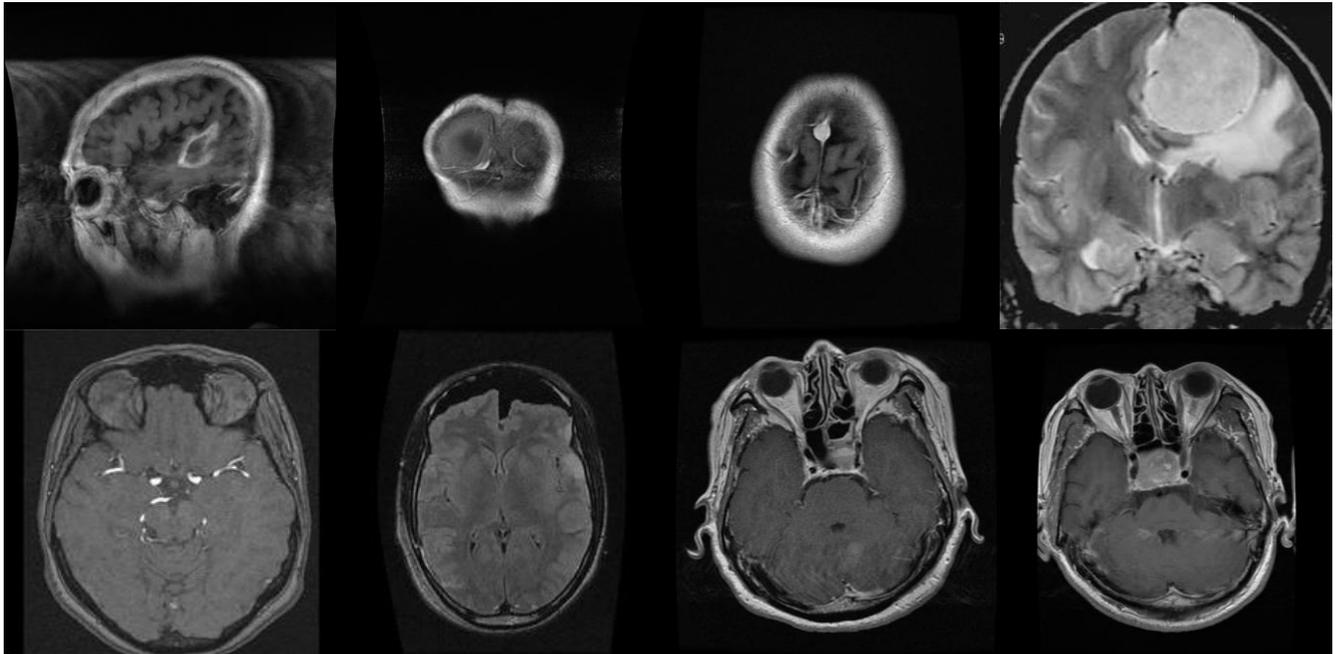

**Figure 5.** Normal and three tumor images those are misclassified by Res-BRNet

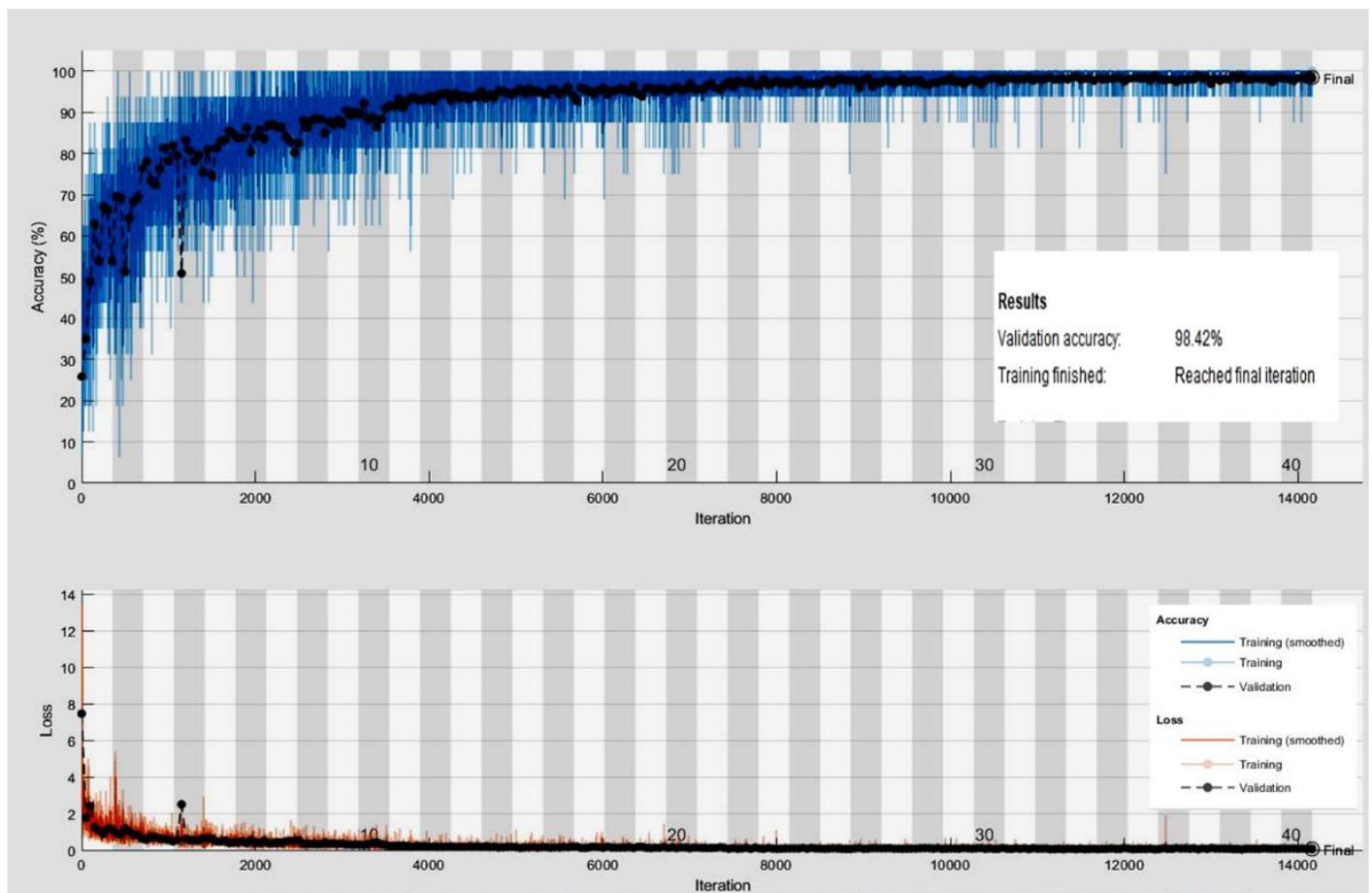

**Figure 6.** Plots of training the proposed Res-BRNet

### 3.2.1. Analysis of performance with base-line methods

The significance of the anticipated idea is assessed by evaluating the performance, especially in comparison to residual learning and spatial exploitation-based architectures. Both baseline architectures, VGG-16 and ResNet-18 are almost deep as Res-BRNet. Spatial block-based architectures exploit one type of down-sampling operation, and residual blocks use stridden convolution instead of pooling down in contrast, employing both pooling operators in Res-BRNet improves the overall performance, as shown in Tab. 1. Thus, according to performance comparison, the Res-BRNet shows exceptional performance compared to residual blocks and Spatial block-based architectures in terms of F1-score (0.9641) and accuracy of (98.22%).

**Table 1.** Performance evaluation of the developed Res-BRNet with baseline architectures using the test data.

| Model | Performance comparison with custom-made CNNs | | | |
|---|---|---|---|---|
| | Accuracy % | Sensitivity | Precision. | F1-Score |
| VGG-16 | 93.32 | 0.9285 | 0.8231 | 0.8719 |
| TL_VGG-16 | 94.66 | 0.9426 | 0.8569 | 0.8961 |
| ResNet-18 | 95.67 | 0.9566 | 0.8788 | 0.9158 |
| TL_ResNet-18 | 96.44 | 0.9641 | 0.8998 | 0.9303 |
| **Proposed Res-BRNet** | **98.22** | **0.9811** | **0.9822** | **0.9641** |

**Fig. 7.** Shows that the suggested Res-BRNet considerably enhances the detection ability for all three brain tumors as well as for normal MRI images compared to baseline residual learning and spatial exploitation architectures.

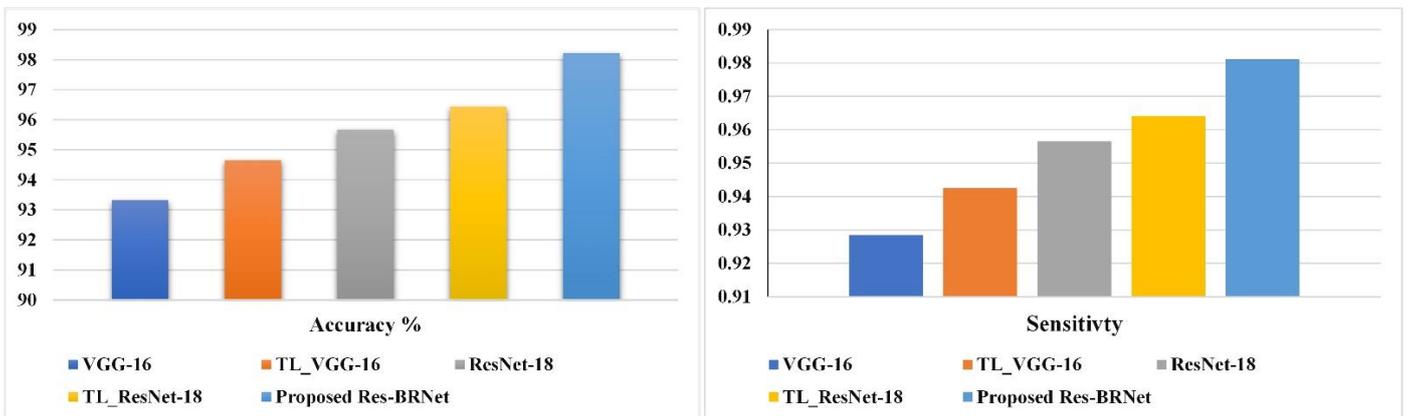

**Figure 7.** Performance assessment of the developed Res-BRNet with baseline architectures

### 3.2.2. Feature space-based Performance Analysis

In order to understand decision-making behaviour, the proposed Res-BRNet and best-performing baseline architectures, ResNet-18 and VGG-16 are evaluated to examine their learned feature spaces.

Characteristics of the feature space responsible for the discrimination capability of a classifier. Features with classes distinguishably improve the model's learning and lower the variance on distinct samples. T-distributed Stochastic Neighbor Embedding (t-SNE) [54] is an algorithm that is well-suited to visualize by embedding high-dimensional points in low dimensions based on similarities between points. Fig. 8 illustrate the 2-D t-SNE plots for the proposed Res-BRNet, ResNet-18 and VGG-16 using testing data. Data visualization shows that the feature space diversity is significantly improved by using both the boundary and regional operations, and it improves the model's performance.

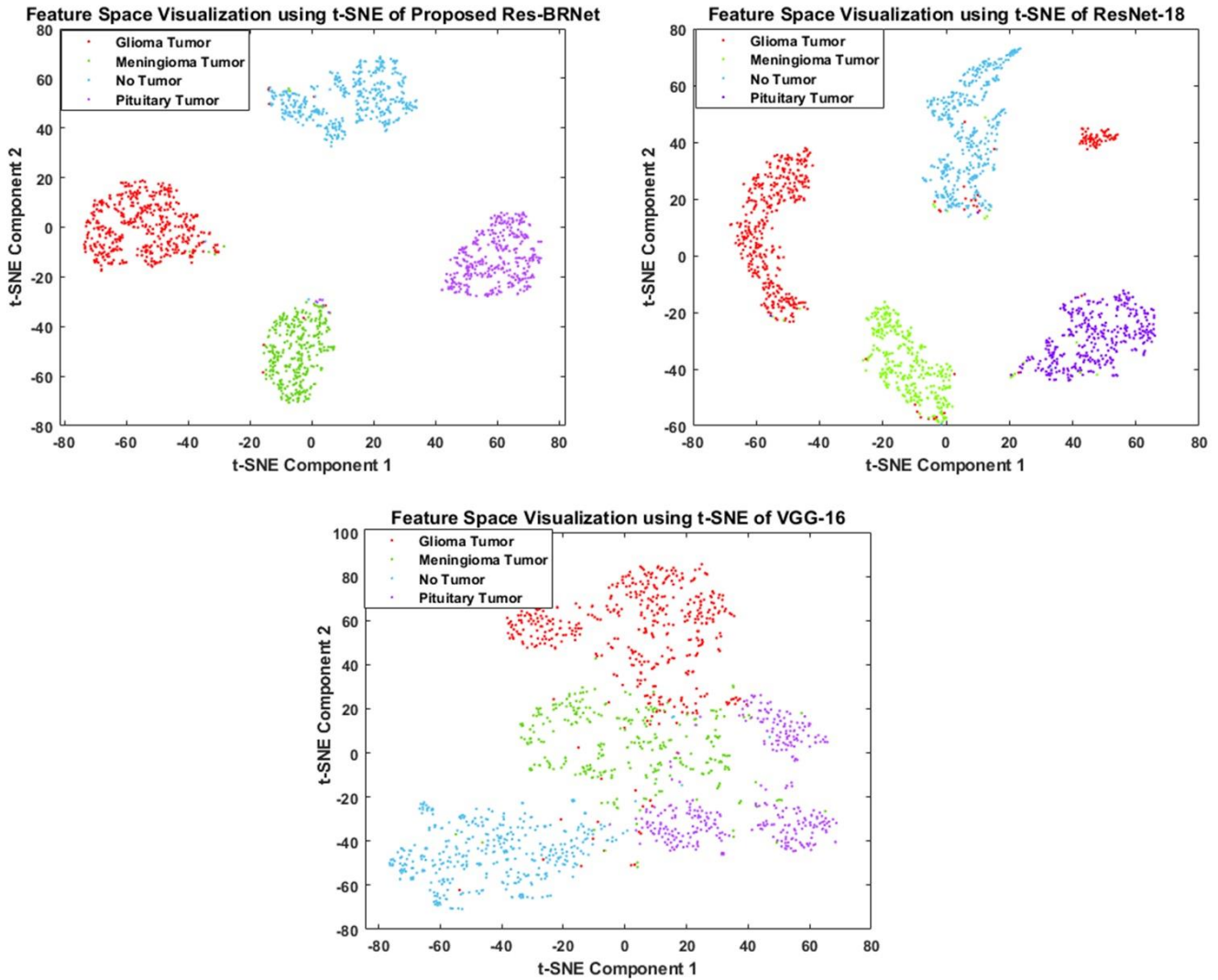

**Figure 8.** Feature space-based performance analysis of the developed Res-BRNet with baseline architectures

### 3.2.3. Performance assessment with existing CNNs

The effectiveness of the developed Res-BRNet is compared with custom-made training-from-scratch and transfer learning-based (TL-based) existing CNNs, namely; SqueezeNet, ShuffleNet, VGG-16, Xception, ResNet-18, GoogleNet, InceptionV3, and DenseNet-201. Tabs. 2 and 3. Illustrate that the proposed models' performance analysis indicates, Res-BRNet is more efficient at identifying the patterns specific to the brain tumors in MRI scans with standard measures of accuracy and F1-score. This performance of the proposed model had improved by using average_ and max_pooling operations systematically in the designed CNN (Fig. 9 and Tab. 4). In general, the model is encouraged to learn highly discriminative features and fine-grained information from the raw MRI image by the use of these opposing pooling operations.

**Table 2.** Performance analysis of the existing standard custom CNNs and the proposed Res-BRNet on the testing data.

| Model | Performance comparison with custom-made CNNs | | | |
|---|---|---|---|---|
| | Accuracy % | Sensitivity | Precision. | F1-Score |
| SqueezeNet | 87.16 | 0.8691 | 0.6946 | 0.7671 |
| ShuffleNet | 89.45 | 0.8923 | 0.7411 | 0.8047 |
| VGG-16 | 93.32 | 0.9285 | 0.8231 | 0.8719 |
| Xception | 95.36 | 0.9531 | 0.8721 | 0.9101 |
| ResNet-18 | 95.67 | 0.9566 | 0.8788 | 0.9158 |
| GoogleNet | 95.87 | 0.9593 | 0.8851 | 0.9196 |
| Inception-V3 | 96.56 | 0.9676 | 0.9015 | 0.9331 |
| DenseNet-201 | 97.01 | 0.9668 | 0.9175 | 0.9406 |
| **Proposed Res-BRNet** | **98.22** | **0.9811** | **0.9822** | **0.9641** |

**Table 3.** Performance analysis of the existing standard TL-based CNNs and the proposed Res-BRNet on the testing data.

| Model | Performance comparison with TL-Based CNNs | | | |
|---|---|---|---|---|
| | Accuracy % | Sensitivity | Precision. | F1-Score |
| TL_SqueezeNet | 90.91 | 0.9108 | 0.7685 | 0.8315 |
| TL_ShuffleNet | 92.31 | 0.9155 | 0.8056 | 0.8521 |
| TL_VGG-16 | 94.66 | 0.9426 | 0.8569 | 0.8961 |
| TL_Xception | 96.37 | 0.9611 | 0.8996 | 0.9285 |
| TL_ResNet-18 | 96.44 | 0.9641 | 0.8998 | 0.9303 |
| TL_GoogleNet | 96.37 | 0.9641 | 0.8985 | 0.9291 |
| TL_Inception-V3 | 97.26 | 0.9711 | 0.9225 | 0.9459 |
| TL_DenseNet-201 | 97.77 | 0.9778 | 0.9349 | 0.9557 |
| **Proposed Res-BRNet** | **98.22** | **0.9811** | **0.9822** | **0.9641** |

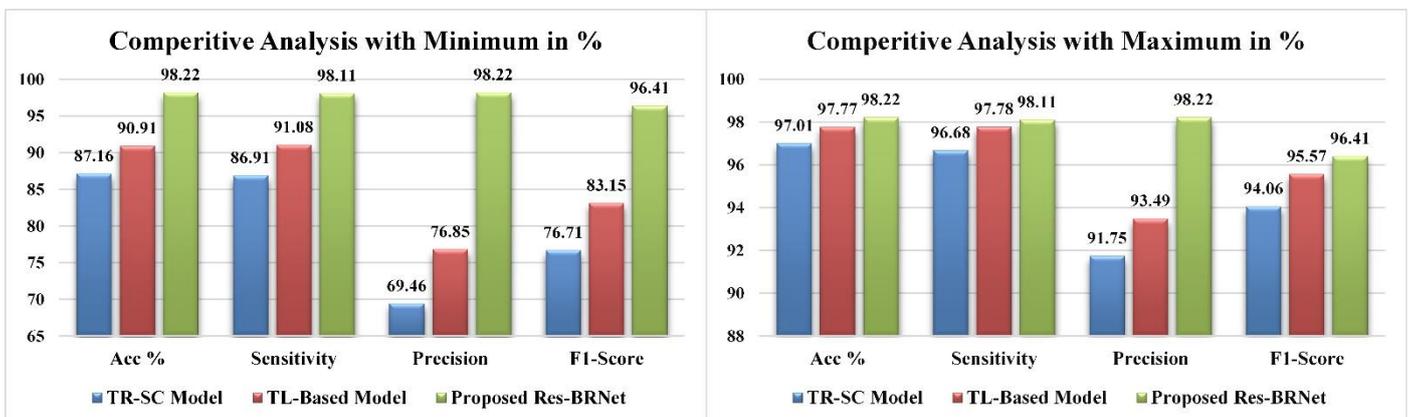

**Figure 9.** Performance improvement of the developed Res-BRNet

**Table 4.** Performance gain of the developed Res-BRNet as compare to TR-SC and TL-Based Models

|   | Improvements | Accuracy % | Sensitivity | Precision | F1-Score |
|---|---|---|---|---|---|
| 1 | TR-SC to TL-B | 0.76 – 3.75 % | 1.1 – 4.45 % | 1.74 – 7.39 % | 1.51 – 6.44 % |
| 2 | TR-SC to Proposed Res-BRNet | 1.2 – 11.06 % | 1.43 -11.02 % | 6.47 – 28.76 % | 2.35 – 19.07 % |
| 3 | TL-B to Proposed Res-BRNet | 0.45 – 7.31 % | 1.1 – 4.17 % | 1.74 – 7.39 % | 1.51 – 6.44 % |

### 3.2.4. ROC and PR-AUC-based analysis

The ROC curve is essential to achieve the optimal analytic threshold for the classifier. ROC curve graphically displays the classifier distinction capability at possible threshold values. As shown in Fig. 10, the proposed Res-BRNet achieved an AUC of (0.9921 and 0.9702) on the brain MRI dataset. ROC and PR-AUC quantitative analysis prove that the suggested method enhances sensitivity by having the lowest false-positive rate. This shows that the presented approach for classifying brain tumors has a lot of potential to be used in the analysis of brain tumors.

### 3.2.5. Screening effectiveness of the proposed technique

Precision and detection rate (sensitivity) are the primary metrics to evaluate a medical diagnostic system's efficiency. The brain tumor detection system needs to have a good detection performance. As can be seen in Fig. 10 and Tab. 2, the detection rate and precision of the proposed approach are evaluated for brain MRI images. As shown in the quantitative study (Fig. 11), the Res-BRNet (Sensitivity: 0.9811, Precision: 0.9822) increases the prediction system's accuracy and has a high prediction rate. Consequently, it is expected to help the radiologist with good accuracy and may be utilized to enhance efficiency by decreasing the burden on medical professionals.

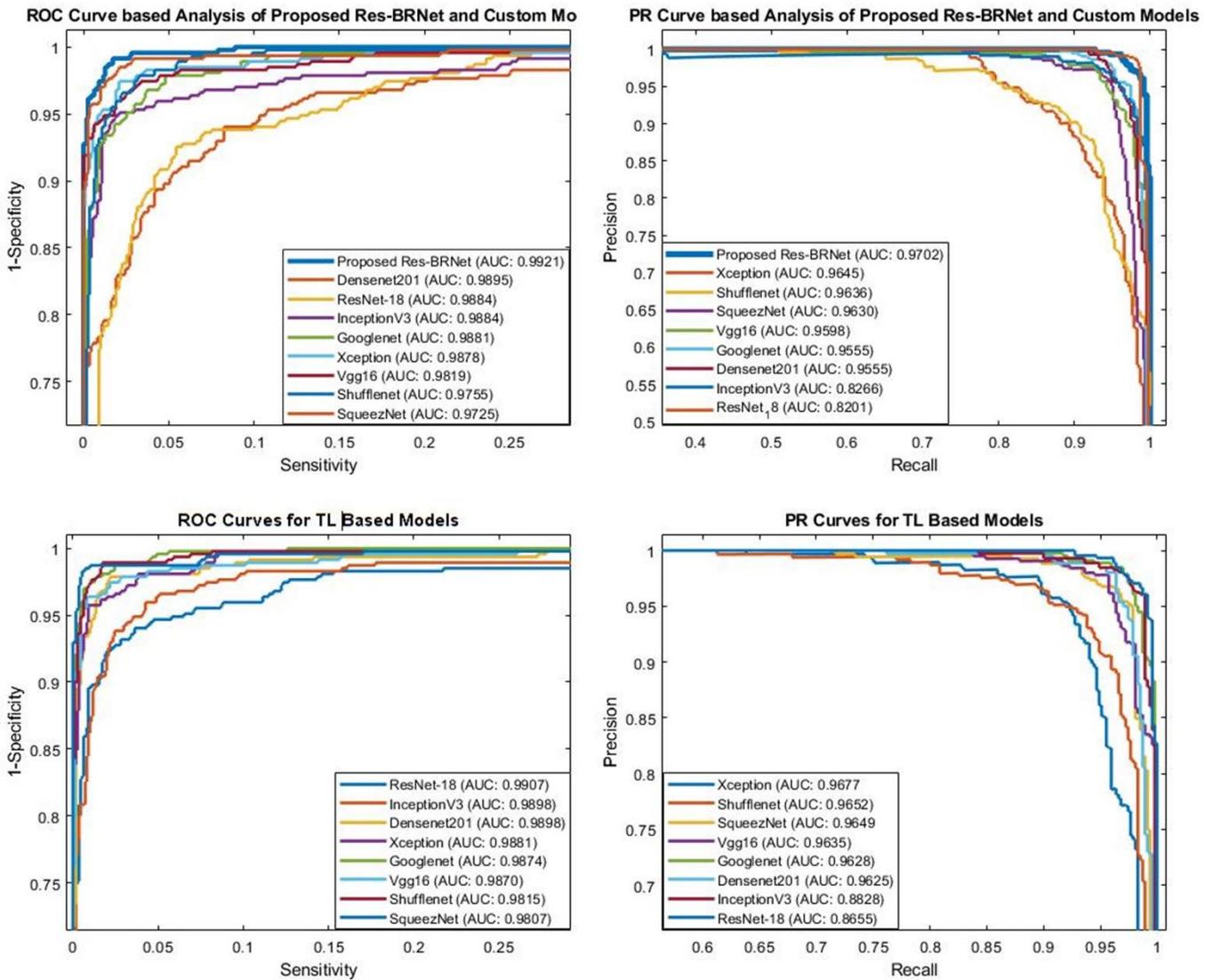

**Figure 10.** Performance analysis of the developed Res-BRNet with existing CNNs

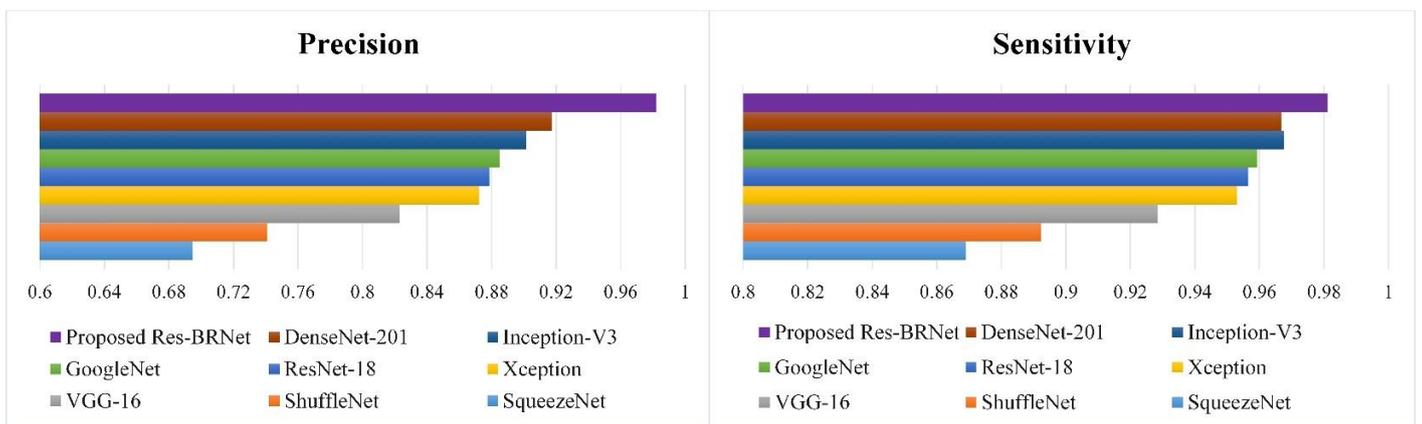

**Figure 11.** Performance analysis of the developed Res-BRNet with existing CNNs

## 4. Conclusion

Brain tumor diagnosis at an early stage is crucial to cure the patient. Therefore, in this work, a new customized deep CNN model is developed to classify the brain MRI scans of meningioma-, glioma-, and pituitary-tumor from healthy entities. The proposed model benefits from data augmentation and learning discriminative features using regional and boundary operators in the developed Res-BRNet. Moreover, the developed Res-BRNet employs spatial and residual ideas to acquire feature-maps with diverse rich information, improving the capability to learn homogeneity, textural variation, and tumor's structural patterns. The performance exploration of the developed is analyzed with the existing Deep CNN models. The experiment results show that the proposed Res-BRNet outperforms existing CNN architectures, indicating accuracy and F1-score improvement. The developed method classifies brain tumors with an accuracy of 98.22%, an F1-Score of 0.9641 with sensitivity and precision of 0.9811 and 0.9822 correspondingly. The proposed approach will likely facilitate healthcare professionals in making diagnoses of brain tumors. Additionally, it motivates us to use in exploring different forms of abnormalities in brain MRI and other medical images.